\documentstyle[epsfig]{elsart}
\def\gapp{\;\raisebox{-.5ex}{$\stackrel{>}{\scriptstyle\sim}$}\;}
\begin{document}
\begin{frontmatter}
\title{D mesons in matter and the in-medium properties of charmonium}

\collab{Bengt Friman${}^a$, Su Houng Lee${}^{a,b,c}$ and Taesoo Song${}^b$}
\address{${}^a$ GSI, Planckstr. 1, D-64291, Darmstadt, Germany \\
${}^b$ Department of Physics and Institute of Physics and Applied Physics,
Yonsei University, Seoul 120-749, Korea \\
${}^c$ Cyclotron Institute, Texas A\&M University, College Station, Texas 
77843, USA}

\begin{abstract}

We study the changes in the partial decay widths of excited
charmonium states into $D \bar{D}$, when the D meson mass decreases
in nuclear matter, taking the internal structure of the hadrons
into account. Calculations within the 3P0 model for $\psi(3686)$
and $\psi(3770)$ imply that naive estimates of the in-medium widths
based only on phase space are grossly exaggerated. Due to nodes in
the wave functions, these states may even become narrow at high
densities, if the D meson mass is decreased by about 200 MeV. For
the $\chi$ states, we generally expect stronger modifications of
the widths. The relevance of the $\chi$ widths for $J/\psi$
suppression in heavy ion collision is discussed. These phenomena
could be explored in experiments at the future accelerator facility
at GSI.

\end{abstract}

}

\end{frontmatter}

\section{Introduction}

The study of hadrons containing heavy quarks is an important tool
for unraveling the properties of the non-perturbative QCD vacuum.
In particular, in mesons consisting of one heavy and one light
quark, such as the D meson, the heavy quark acts as an almost
static source for the light quark, which in turn probes the QCD
vacuum. This is the basic constituent quark picture of heavy-light
meson systems within the formulation based on heavy quark
symmetry~\cite{Wisgur}. Therefore, changes in the vacuum
condensates, e.g., at finite temperature and density, are expected
to affect the light quark and consequently also to modify the
in-medium properties of the $D$ meson. This is explicitly borne out
in the model calculations based on QCD sum rule
analysis~\cite{Hayashigaki00,Morath} and the Quark Meson coupling
model~\cite{Tsushima99}, both of which show a reduction of the D
meson mass of about 50 MeV at nuclear matter density. Lattice gauge
theory results for the heavy quark potential at finite temperature,
suggests a similar drop of the D meson mass at finite
temperature~\cite{Digal01}.

The reduction of the $D$-meson mass obviously has important
phenomenological consequences, ranging from the possible existence
of charmed mesic nuclei~\cite{Tsushima99} to enhanced subthreshold
production of open charm in $\bar{p}A$
reactions~\cite{Sibirtsev99}. Moreover, a reduction of the D-meson
mass in matter has also direct consequences for the production of
open charm~\cite{Cassing01} and the suppression of
$J/\psi$-mesons~\cite{Sibirtsev00} in relativistic heavy ion
collisions. Such phenomena could be explored in experiments e.g. at
the future accelerator facility at GSI~\cite{GSI-future}.

The decrease of the D-meson mass in matter is by and large due to
the restoration of chiral symmetry in the nuclear medium. This is
demonstrated by QCD sum rule calculations of the D meson mass in
vacuum~\cite{Aliev83} and in matter~\cite{Hayashigaki00,Morath}. In
the D-meson sum rules, the leading term is dominated by chiral
symmetry breaking operators. This is contrary to heavy quarkonia,
such as $J/\psi$ and $\psi(3686)$, where the constituent quark and
antiquark probe only the gluonic condensates. Consequently, such
states are expected to experience only a relatively small mass
shift in medium~\cite{Bro90,Was91,Luk92,Kli99,Hay99,KL01}. This
implies that, with increasing density, the $D \bar{D}$
threshold~\cite{Hayashigaki00,Digal01,Wong01,Wetal} will cross the
energy of some of the excited charmonium states. Consequently, the
reduction of the $D\bar{D}$ threshold in matter may be reflected in
the dilepton spectrum of $\bar{p} A$ or $A A $ reactions as an
increased width of the peaks corresponding to the $\psi(3686)$
state. Furthermore, since more than 40\% of the $J/\psi$ produced
in heavy ion collision emanate from the $\psi(3686)$ and $\chi$
states~\cite{chi}, such crossings will induce a stepwise
suppression of $J/\psi$ signal due to the successive melting of
excited charmonium states at finite density and
temperature~\cite{Hayashigaki00,Digal01,Wong01,Wetal}.

In this letter, we point out that the level crossing between the
charmonium and the $D \bar{D}$ threshold does not result in an
immediate dissolving of the charmonium states, as found in naive
calculations, where the participating mesons are effectively
treated as point particles. When the internal structure is taken
into account, the effective coupling to the $D\bar{D}$ final state
depends strongly on the Q-value, and consequently on the momentum
carried by $D$ meson in the charmonium rest frame. The overlap of
the $D$ mesons with the wavefunction of the initial heavy quarks of
the charmonium depends strongly on the relative momentum of the $D$
mesons in the final state. In other words, the effective $\psi
D\bar{D}$ coupling constant is sensitive to the wave functions and
the momenta of the particles involved in the decay. This mechanism
provides a viable interpretation~\cite{LeY77} of the experimentally
observed branching ratios for the decay of the $\psi(4040)$ into $D
\bar{D}, D \bar{D}^*, D^* \bar{D}^*$. 
There are strong deviations from predictions
based on naive quark spin counting. It was shown that the effective
coupling between $\psi(4040)$ and the $D$ mesons varies rapidly
with the relative momentum in the final state. The matrix elements
may even vanish at certain momenta, corresponding to nodes in the
wave function~\cite{LeY77}. This mechanism was confirmed also in a
more sophisticated potential model for the heavy
quarkonium~\cite{cornel}.

As we will show, a similar mechanism is active also in the decay of
$\psi(3686)$, $\psi(3770)$, $\chi_{c0}(3417)$ and $\chi_{c2}(3556)$
into $D \bar{D}$ as the mass of the D meson decreases.  Due to the
nodes in the radial ($\psi(3686)$) and orbital ($\psi(3770)$) wave
function, the partial decay widths first increase and then decrease
as the mass of the D meson is reduced. In particular we find that,
for a mass shift of $200-250$ MeV, the branching ratio into the
$D\bar{D}$ channel vanishes and then increases again when the $D$
mass is reduced further. The resulting widths are much smaller than
those obtained in the naive picture, where the width is enhanced
due to the increase in phase space, while the coupling constant is
kept fixed.

For the $\chi$ mesons the picture is somewhat different. The
partial width of the $\chi_{c0}(3417)$, increases very rapidly as
the $D\bar{D}$ channel opens up, and then approaches zero as the
mass of the D meson is decreased even further. On the other hand,
for the $\chi_{c2}(3556)$ the partial decay width increases slowly
and monotonically, because there is no node in the wave function.

\section{Charmonium states}

We use the harmonic oscillator potential to model the bound state
wave functions and the 3P0 model to describe the charmonium decays.
In this exploratory calculation our aim is to determine the
in-medium properties of the excited charmonium states on a
semi-quantitative level. Our main result, the medium dependence of
the effective coupling constants, is due mainly to the node
structure of the wave functions and does not depend strongly on the
details of the model. Hence, a more sophisticated calculation based
on a refined potential will not change the main conclusions of our
work. In the harmonic oscillator potential model, the wave function
of a heavy quarkonium state is of the form,
\begin{eqnarray}
\phi_{N,l} = {\rm Normalization} \times Y_l^m(\theta, \phi)
(\beta^2 r^2)^{\frac{1}{2}l} e^{- \frac{1}{2} \beta^2 r^2}
L^{l+\frac{1}{2}}_{N-1}
(\beta^2 r^2)
\label{wavefunction}
\end{eqnarray}
where $\beta^2= M \omega/\hbar$ characterizes the strength of the
harmonic potential, $ M= \frac{1}{2} m_{c}$ is the reduced mass of the 
charm quark anti-quark  system, and
$L_p^k(z)$ is a Laguerre Polynomial.  
The energy eigenvalues are
\begin{eqnarray}
E_n=\hbar \omega (n+\frac{1}{2}), ~~~~n=2k+l+1=2(N-1)+l+1,
\end{eqnarray}
where, $N$ is the number or nodes in the radial direction,
including one at infinity. In table I, we summarize the quark-model
assignments of quantum numbers and other relevant information on
the lowest lying charmonium states. For the $\psi(3770)$, the
coupling constant $g_{\psi DD}$ is determined by fitting its width
using $\Gamma_{Tot}=\Gamma_{\psi\rightarrow DD} = (g_{\psi DD}^2/24
\pi) ( (m_\psi^2-4m_D^2)^{3/2}/ m_\psi^2)$. For the $J/\psi(3097)$,
which is far below the $D\bar{D}$ threshold, we use vector meson
dominance for the electromagnetic current of the $D$-meson,
assuming that the form factor is dominated by the $J/\psi$ at small
$q^2$. In vacuum the $D\bar{D}$ threshold is located at 3.74 GeV.

\begin{table}
\caption{Charmonium states in the harmonic oscillator model.}
\vspace*{0.2cm}
\begin{center}
\begin{tabular}{cccccc} \hline
Charmonium  &  $N\,^{2S+1} L_J$ & HO energy  & $\Gamma(e^+e^-)$
(KeV) & $\Gamma(Tot)$ (MeV) & $g_{\psi DD}$    \\
\hline\hline
 $ J/\psi(3097)$ & $1\,^3S_1$ &  $\frac{3}{2} \hbar \omega$   &
5.26 & 0.087 &    7.8  \\
 $\psi(3686)$ & $2\,^3S_1$ &  $(\frac{3}{2}+2) \hbar \omega$ &
 2.14 & 0.277 &
       \\ $\psi(3770)$ & $1\,^3D_1$ &  $(\frac{3}{2}+2) \hbar
 \omega$ & 0.28 &  23.6  & 15.4   \\
\hline
$\chi_{c0}(3417)$ & $1\,^3P_0$ & $(\frac{3}{2}+1) \hbar \omega$ & &
14 &   \\
 $\chi_{c1}(3510)$ & $1\,^3P_1$ & $(\frac{3}{2}+1) \hbar \omega$ &
 & 0.88  &  \\ $\chi_{c2}(3556)$ & $1\,^3P_2$ & $(\frac{3}{2}+1)
 \hbar \omega$ &    &  2  &  \\
\hline
\end{tabular}
\end{center}
\label{tab1}

\end{table}

\begin{figure}
\begin{center}
\includegraphics[height=4cm]{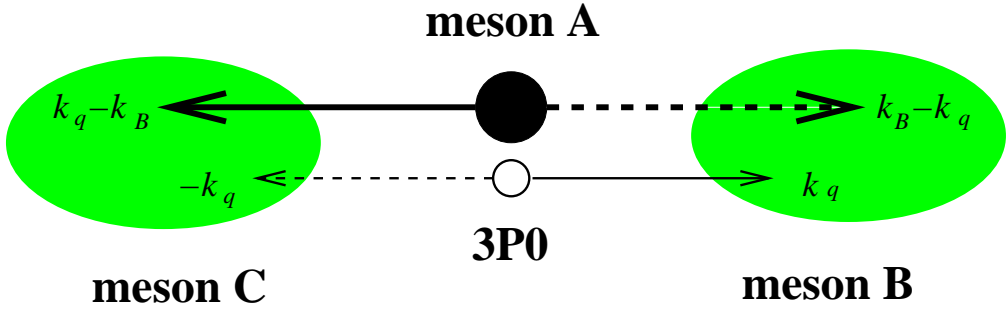}
\caption{\label{fig:fig1} Diagrammatic representation of meson A decaying
into mesons B and C by creating a 3P0 quark anti-quark pair.}
\end{center}
\end{figure}

\section{The 3P0 model}

We compute the decay of the charmonium states into $D \bar{D}$
mesons using the 3P0 model~\cite{LeY73}. In this model, the decay
of a meson A into mesons B and C involves the following invariant
matrix element
\begin{eqnarray}
M_{A \rightarrow BC}= \langle A| \gamma \left[\bar{q}_s
q_s\right]^{{}^3P_0} |BC \rangle,
\label{matrix}
\end{eqnarray}
where $\gamma$ is a coupling strength, which characterizes the
probability for creating a quark anti-quark pair in the $3P_0$
state. In fig.~1 we show a diagrammatic representation in the rest
frame of the meson $A$ consisting of a heavy quark anti-quark pair
with momenta $k_q-k_B$ and $k_B-k_q$. The decay products are the
mesons $B$ and $C$ with total momenta $k_B$ and $-k_B$ respectively.
The meson $B$ is composed of a heavy anti-quark (momentum $k_B-k_q$)
emanating from the parent meson $A$ and a quark (momentum $k_q$) from
the 3P0 pair. Now, since the decaying quarks has to form the
physical meson, the matrix element will involve the following
overlap integral,
\begin{eqnarray}
M_{A \rightarrow BC}&  \propto &
\int d^3k_q  \phi_A(2k_q-2k_B) \phi_B(2k_q-k_B) \phi_C(2k_q-k_B)
\nonumber \\
&&  ~~~~~~ \times [\bar{u}_{k_q,s} v_{-k_q,s}]^{{}^3P_0}
\label{overlap}
\end{eqnarray}
where the mesonic wave functions $\phi$ are the Fourier transforms
of the spatial wave functions in Eq.~(\ref{wavefunction}). The
momentum space wave functions are polynomial functions in the
relative momentum multiplied by gaussians. Finally, $[\bar{u}_{k_q,
s} v_{-k_q,s}]^{{}^3P_0}\propto k_qY^m_1(k_q)$ is the wave function
of the produced $\bar{q} q$ pair.

The overlap integral in eq.(\ref{overlap}), is a function of the
meson momentum $k_B$. When there are nodes in one of the wave
functions $\phi$, the integral can vanish for certain values of
$k_B$. This is seen by combining the three gaussians in
eq.~(\ref{overlap}) into one. To this end one makes a change of
variables
\begin{eqnarray}
k_q^\prime = k_q -{ 1+ r^2 \over 1 +2 r^2} k_B,
\label{change}
\end{eqnarray}
where $r={\delta \over \beta}$, 
$\delta$ is the strength of the  HO potential of the parent   
meson $A$ and $\beta$ is that of the emitted mesons $B$ and $C$.
We introduce two independent parameters $\beta$ and $\delta$ 
to allow for different sizes of the wave function for the 
initial meson and the outgoing mesons.   
Now, if one or more of the mesons involved is in an excited state,
the polynomial in the corresponding wave function $\phi$ is, after
the change of variables (\ref{change}), a polynomial in $k_B$ and
$k_q^\prime$. Consequently, the matrix element (\ref{overlap}) is
proportional to a polynomial in $k_B$, which may vanish for certain
values of $k_B$.

For the case of interest, where the charmonium decays into two
pseudoscalar mesons $D\bar{D}$, the relevant formulae in the  3P0
model can be found in ref.~\cite{Barnes97} for  $\beta=\delta$. 
Here, we re-derive the formula in 3P0 model, allowing for  
$\beta \neq \delta$.  
The invariant matrix
element for the decay $A\rightarrow B+C$ is given by,
\begin{eqnarray}
M_{LS}= {\gamma \over \pi^{1/4} \beta^{1/2}} {\cal P}_{LS}(x,r)
e^{-\frac{x^2}{4(1+2 r^2)}} \times \frac{1}{2} [I(d_1)+I(d_2)].
\end{eqnarray}
In our case, the flavor factors $\frac{1}{2} [I(d_1)+I(d_2)] =
\frac{1}{2}$. Furthermore,
\begin{eqnarray}
x=\frac{1}{\beta} \times \sqrt{m_A^2/4-m_B^2}\,,
\label{defx}
\end{eqnarray}
which is the scaled momentum carried by the decaying mesons in the
rest frame of the parent meson $A$. The decay rate is then given by
\begin{eqnarray}
\Gamma (A\rightarrow B+C)=
2\pi {p_B E_B E_C \over M_A} \sum_{LS} |M_{LS}|^2\,.
\end{eqnarray}

We now present the resulting decay rates for the different
charmonium states. The quark-model assignments of these states are
given in Table 1.  For $r=1$ $(\beta=\delta)$,  
our results reduce to those given of Barnes {\em et al.}~\cite{Barnes97}.
\begin{enumerate}

\item $\chi(3417)$
\begin{eqnarray}
{\cal P}_{00}^{(1~{}^3P_0 \rightarrow {}^1S_0+{}^1S_0)}
=\sqrt{\frac{3}{2}} \times  2^5 \bigg( { r \over 1+2 r^2} \bigg)^{5/2}
\biggl( 1-{ (1+r^2) \over 3 (1+2 r^2)} x^2 \biggr),
\end{eqnarray}
\begin{eqnarray}
\Gamma(\chi(3417) \rightarrow D +\bar{D})
& = & {\pi^{1/2} E_D^2 \over M_{\psi(3417)}} \gamma^2
2^{9}  3 \biggl( { r \over 1+2r^2} \biggr)^5 x \nonumber \\
&& \times \biggl( 1-{ (1+r^2) \over 3 (1+2 r^2)} x^2 \biggr)^2 
e^{-\frac{x^2}{2 (1+2r^2)}}.
\label{g3417}
\end{eqnarray}

\item $\chi(3556)$
\begin{eqnarray}
{\cal P}_{20}^{(1~{}^3P_2 \rightarrow {}^1S_0+{}^1S_0)}
=\frac{1}{\sqrt{15}} \times
{ r^{5/2} 2^5 (1+r^2)  \over (1+2 r^2)^{7/2}} x^2,
\end{eqnarray}
\begin{eqnarray}
\Gamma(\chi(3556) \rightarrow D +\bar{D})
={\pi^{1/2} E_D^2 \over M_{\psi(3556)}} \gamma^2
{2^{10}  \over 15} 
{ r^5 (1+r^2)^2  \over (1+2 r^2)^7}
x^5 e^{-\frac{x^2}{2 (1+2r^2)}}.
\label{g3556}
\end{eqnarray}

\item $\psi(3686)$
\begin{eqnarray}
{\cal P}_{10}^{(2~{}^3S_1 \rightarrow {}^1S_0+{}^1S_0)}
& =&{ 2^{7/2} (1-3r^2)(3+2r^2) \over 3 (1+2r^2)^{7/2}} x \nonumber \\ 
&& \biggl(1+{ 2 r^2(1+r^2) \over (1+2r^2) (3+2r^2)(1-3r^2)}x^2 \biggr),
\end{eqnarray}
\begin{eqnarray}
\Gamma(\psi(3686) \rightarrow D +\bar{D})
 =  {\pi^{1/2} E_D^2 \over M_{\psi(3686)}} \gamma^2
{2^{7}  \over 3^2} { (3+2r^2)^2 (1-3r^2)^2 \over (1+2r^2)^7} 
x^3  \nonumber \\
 \times
\biggl(1+{ 2 r^2(1+r^2) \over (1+2r^2) (3+2r^2)(1-3r^2)}x^2 \biggr)^2
e^{-\frac{x^2}{2(1+2r^2)}}.
\label{g3686} 
\end{eqnarray}

\item $\psi(3770)$
\begin{eqnarray}
{\cal P}_{10}^{(1~{}^3D_1 \rightarrow {}^1S_0+{}^1S_0)}
=-{2^5 \sqrt{10}  \over 3}
\biggl( {r \over 1+2 r^2} \biggr)^{7/2}
x \biggl(1-{1+r^2 \over 5(1+2 r^2)} x^2 \biggr),
\end{eqnarray}
\begin{eqnarray}
\Gamma(\psi(3770) \rightarrow D +\bar{D})
& = & {\pi^{1/2} E_D^2 \over M_{\psi(3770)}} \gamma^2
{2^{11} 5 \over 3^2}
\biggl( {r \over 1+2 r^2} \biggr)^7
 x^3 \nonumber \\
& & \times
\biggl(1-{1+r^2 \over 5(1+2 r^2)} x^2 \biggr)^2
e^{-\frac{x^2}{2(1+2r^2)}}.
\label{g3770}
\end{eqnarray}
\end{enumerate}
Note that in each case, $x$ is defined by eq.(\ref{defx}) with
$m_A$ the mass of the corresponding charmonium state and $m_B=m_D$.
The zero in eq.(\ref{g3686}) is due to the nodes in the radial wave
function, whereas those in eq.~(\ref{g3417}) and eq.~(\ref{g3770})
results from the orbital part. Nevertheless, for $r=1$, the two widths in
eq.(\ref{g3686}) and eq.(\ref{g3770}) have the same functional
form.

Our model has three parameters, namely, $\beta$, $r=\delta/\beta$ and $\gamma$:

\begin{itemize}
\item $\beta$ determines the size of the harmonic oscillator potential 
of the $D$ mesons, while $\delta$ is the corresponding parameter for charmonium.  
We fix $\beta$ and the ratio $r$ by fitting the partial decay width of $\psi(4040)$
to $D D$, $D D^*$ and $D^*D^*$ to the experimental ratios of 1:20:640, 
as has been done in ref.~\cite{LeY77} for $r=1$. 
The formulas of ref.~\cite{LeY77} can be generalized to $r\neq 1$ by the following 
replacement,
\begin{eqnarray}
\frac{35}{4\cdot 3^3}
 \biggl(  1- \frac{4}{15}x^2+\frac{4}{315}x^4 \biggr)
  \to   \biggl( 
\frac{15}{8} {1+r^2 \over 1+2r^2}- { 5 r^2(4+r^2) \over(1+2r^2)^3}
\nonumber \\
+{r^2(5-9r^2-10r^4) \over 2 (1+2r^2)^4}x^2 
+ {r^4(1+r^2) \over 2(1+2r^2)^5 } x^4 \biggr)
\end{eqnarray}
The fit yields,
\begin{eqnarray}
\beta=  0.3\,\,  {\rm GeV},~~~{\rm and}~~~r=1.04\,.
\label{beta1}
\end{eqnarray}
The resulting oscillator parameter $\beta$ is close to the value obtained in
ref.~\cite{LeY77} ($\beta=0.31$ GeV) and the ratio $r$ is close to unity. The
implied difference in size of the $D$ meson and charmonium is very small and,
in view of the simplicity of the model, presumably insignificant.  Within the
harmonic oscillator model, one can also relate $\beta$ to the mass splitting
between the charmonium states if one assumes an effective charm quark mass
$m_c$.  In this case, for $m_c = 1.5$ GeV, we find $\beta \sim$ 0.43 GeV.
However, since we are interested in the partial decay widths of charmonium
states, we will choose the value given in eq.(\ref{beta1}).

\item $\gamma$ determines the strength of the 3P0 vertex.
This parameter may be determined by fitting the
$\Gamma(\psi(3770)\rightarrow D
\bar{D})$, which implies $\gamma =0.281$.
\end{itemize}

These values for the parameters are qualitatively consistent with
those obtained from a best fit to the decays of excited states of
light mesons~\cite{Barnes97}, $\beta=0.36$ GeV and $\gamma=0.4$.

\begin{figure}
\begin{center}
\includegraphics[height=8cm,angle=-90]{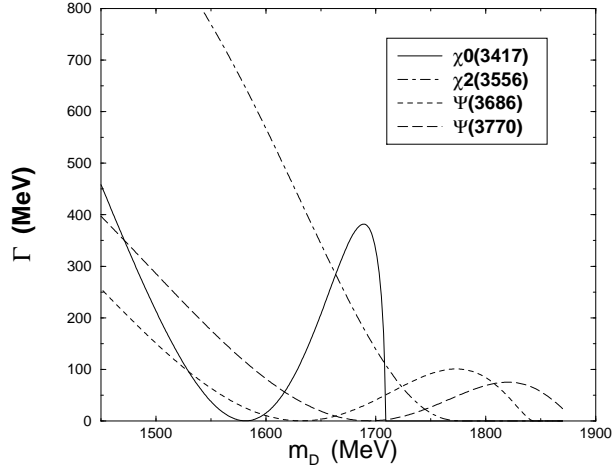}
\caption{\label{fig:fig2} Partial decay width of charmonium decaying into
$D \bar{D}$, as a function of the D meson mass.}
\end{center}
\end{figure}
In fig.~2 we show the dependence of the partial widths for the
decay into the $D\bar{D}$ channel on the D meson mass in medium.
For the $J^{PC}=1^{--}$ states, we note that when the $D$-meson
mass is reduced, $\Gamma_{D\bar{D}}(\psi(3770))$ and $\Gamma_{D
\bar{D}}(\psi(3686))$ first grow to about 90 MeV for a mass
shift of about 50 MeV. Due to the nodal structure of the wave
function, a further reduction of the $D$-meson mass leads to a
strong decrease of the partial width, which then vanishes for a
mass shift of about 200 MeV and 250 MeV, respectively. In both
cases, the increase of the width beyond this point is slow, so that
even when the D meson mass is reduced by 400 MeV, the widths remain
below 350 MeV. In fig.~3, we illustrate the important role of the
internal structure of the mesons by comparing the in-medium width
of the $\psi(3770)$ (eq.(\ref{g3770})) with the naive estimate
obtained by treating the mesons as point particles
\begin{eqnarray}
\Gamma_{D \bar{D}}^\star(\psi(3770)) = \frac{g_{\psi DD}^2}{24 \pi}
{ (m_\psi^2-4m_D^2)^{3/2} \over m_\psi^2}.
\end{eqnarray}
Here the coupling constant is kept fixed at its vacuum value
$g_{\psi DD}=15.4$. Consequently,
$\Gamma_{D\bar{D}}^\star(\psi(3770))$ grows rapidly with dropping
$D$-meson mass, due to the strong increase in available phase
space. Clearly, the width of the $\psi(3770)$ in matter is strongly
overestimated in the naive calculation, where the internal meson
structure is ignored. Consequently, the expectation that the $\psi$
states would melt instantly as the $D\bar{D}$ channel opens up is
not well founded.

As shown in fig.~2, the situation for the $\chi$ states is somewhat
different. The width of the lightest $\chi$ meson , the
$\chi_{c0}$, increases very rapidly as the decay into the
$D\bar{D}$ channel becomes possible. Again, a node in the wave
function leads to a zero in the width as the $D$-meson mass is
decreased further. The sudden increase at threshold is due to the
fact that in $J^\pi=0^+$, the $D\bar{D}$ pair is in an s-wave
state. This implies, that $\chi_{c0}(3417)$ will dissolve, when the
$D\bar{D}$ channel opens up. However, this will probably have only
little impact on $J/\psi$ suppression in heavy ion collisions,
since the feeding of $J/\psi$'s from the decay of the
$\chi_{c0}(3417)$ is expected to be negligible~\cite{chi}.

On the other hand, the $D\bar{D}$ width of the $\chi_{c2}(3556)$ is
suppressed at first, because for $J^\pi=2^+$ the $D$ mesons are in
an $l=2$ state. As the $D$-meson mass is decreased further, this
partial width increases monotonically, because the wave function of
this state has no nodes. Thus, for large mass shifts, its width is
much larger than for the other charmonium states. This behavior
could have an important effect on $J/\psi$ suppression in heavy ion
collision, since at sufficiently high densities, the produced
$\chi_{c2}(3556)$ mesons will melt instantly and not contribute to
the production of $J/\psi$'s in heavy ion collision.  This would
eliminate more than 10\% of the expected $J/\psi$ yield~\cite{chi}.

\begin{figure}
\begin{center}
\includegraphics[height=8cm,angle=-90]{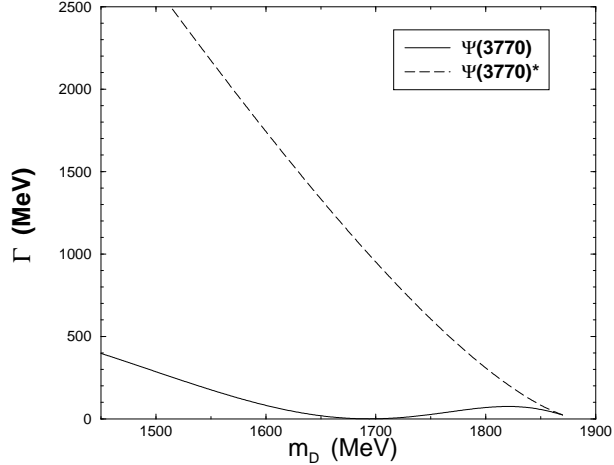}
\caption{\label{fig:fig3} The solid line is the partial decay width
 $\Gamma(\psi(3770) \rightarrow D + \bar{D})$ from 3P0 model.  The
dashed line is assuming a constant coupling normalized to give the
same leading behaviour at the threshold point.}
\end{center}
\end{figure}

\section{Conclusion}

We have shown that the internal structure of the mesons play a
crucial role in calculations of the in-medium widths of charmonium
states. In particular, we find that the dropping of the $D$ meson
mass in nuclear matter does not lead to the extreme increase of the
decay width of charmonium states, expected from naive phase space
arguments. In fact, for both $\psi(3686)$ and $\psi(3770)$ we find
that the partial decay width vanishes for a particular D meson mass
shift of about 200 to 250 MeV, due to cancellations in the matrix
elements caused by the nodal structure of the quark wave functions.
For a mass shift of this size, the increase in phase space, would
give $\Gamma_{D\bar{D}}^\star(\psi(3770))\sim 1.2$ GeV for a
constant matrix element. The enhanced widths of the $\psi(3686)$
and $\psi(3770)$ may be observable in the spectrum of dileptons
produced in heavy ion reactions. As can be seen in fig.~2, the
partial width for decay into the $D\bar{D}$ channel first increases
to about 90 MeV and then drops to zero for a D meson mass shift of
about 200 MeV.

In matter the total width of charmonium states is further enhanced
by reactions with the surrounding particles, like e.g. $\psi
+N\rightarrow\Lambda_c+\bar{D}$. The threshold for this process is
fairly low; it is energetically allowed for charmonium masses
$m_\psi \gapp 3.2$ GeV. The cross section, computed in the quark
exchange model, is given in ref.~\cite{Martins96}. For the
$\psi(3686)$ the asymptotic value was found to be
$\sigma_{abs}(\psi(3686)+N)\sim 6$ mb. The relatively small cross
section is related to the compact size of the charmonium states.
The contribution of this reaction to the in-medium width of the
$\psi(3686)$ is, to lowest order in density, given by
\begin{eqnarray}
\Gamma_N=\frac{1}{\tau} =\langle \sigma_{abs}(\psi(3686)+N)
v_{rel} \rho_n \rangle,
\end{eqnarray}
where $\rho_n$ is the density of nucleons in nuclear matter and
$v_{rel}$ is the average relative velocity in the initial state.
Assuming that the $\psi$ has a relative momentum $p_\psi$ with
respect to the rest frame of the medium , $v_{rel}=\frac{3}{4}
\frac{p_F}{m_N}[1+\frac{2}{3}(\frac{m_N p_\psi}{m_\psi p_F})^2]$,
where $p_F$ is the Fermi momentum of nuclear matter. When
$p_\psi=0$, this gives the additional width $\Gamma_N \sim 4$ MeV
at $\rho_n=\rho_0=0.17 $fm$^{-3}$ and $\Gamma_N \sim$ 25 MeV at
$\rho_n=4\times \rho_0$. Even for $p_\psi=1$ GeV$/c$, we find an
additional width of only $\Gamma_N \sim 6.9$ MeV at $\rho_n=\rho_0$
and $\Gamma_N \sim$ 34 MeV at $\rho_n=4\times \rho_0$.

Although there are no calculations of the corresponding reaction
for the $\psi(3770)$, one expects a cross section of the same
magnitude because the root-mean-square radii of the two charmonium
states are similar~\cite{cornel}. In QCD~\cite{Peskin79,Lee02}, 
the leading order result for $\sigma_{abs}$ is proportional to
$\langle r^2\rangle$. Therefore, the additional width of the
charmonium states $\psi(3686)$ and $\psi(3770)$ due to scattering
off the nuclear medium is expected to be small. Thus, we conclude
that the total width of these resonances in nuclear matter is
dominated by the decay into $D\bar{D}$ mesons. The small scattering
contribution may play a significant role only at those densities,
where the $D\bar{D}$ channel is quenched due to the nodal structure
of the wave function.

For the $\chi(3556)$ a somewhat different picture emerges. We find 
that its width increases monotonically as the D meson mass decreases.
In hadronic collisions~\cite{chi} a large fraction of the $J/\psi$'s stem from
the radiative decay of $\chi$'s. If the $D\bar{D}$ width of the $\chi(3556)$
increases in matter, the probability for the decay into $J/\psi \gamma$
decreases. Thus, part of the suppression of $J/\psi$'s in heavy-ion collisions
may be due to this effect. The increase in the width of the $\chi(3556)$ could
be observed by measuring the $J/\psi \gamma$ decay \cite{chi} in future heavy
ion experiments~\cite{GSI-future}.

In addition to the medium effects due to the partial restoration of
chiral symmetry, one also expects mass changes due to modifications
of the confining potential in nuclear matter. In our model, this
effect will be reflected in a density dependence of the oscillator 
parameters $\beta$ and $\delta$, which
parameterize the strength of the harmonic oscillator potential. A
rough estimate of the expected change in $\delta$ at nuclear matter
density, can be obtained by relating the expected mass shift of the
$J/\psi$ to a change in the confining potential. We use the
relation $\delta m_{J/\psi}=\frac{3}{2} \frac{1}{M}
\delta p^2$, which applies to the harmonic oscillator model. All
model calculations~\cite{Bro90,Was91,Luk92,Kli99,Hay99,KL01} yield
$\delta m_{J/\psi} \sim $ -7 MeV at nuclear matter density. This
translates into $\delta \beta = -5.6 $MeV, which can be safely
neglected, since it corresponds to a relative shift in $\beta$ of less than
2 \% (see eq.~(\ref{beta1})) and an even smaller change in the final 
result in fig.~2.

We conclude that measurements of the in-medium properties of
charmonium states can provide valuable information on the
characteristics of QCD in dense and hot matter.

\section{Acknowledgement}

We thank D.-O. Riska for useful comments. SHL gratefully
acknowledges hospitality of GSI, where this work was initiated and
completed. The work was supported in part by the KOSEF Grant
1999-2-111-005-5, and the Korean Ministry of Education Grant
2000-2-0689.

\end{document}